 \documentclass[preprint2]{aastex}

\usepackage{lscape}

\shorttitle{Unusual quadruple system HD 91962}
\shortauthors{Tokovinin, Latham, \& Mason}
\begin{document}

\renewcommand{\topfraction}{1.0}
\renewcommand{\bottomfraction}{1.0}
\renewcommand{\textfraction}{0.0}

\title{The unusual quadruple system HD 91962 with a ``planetary'' architecture}


\author{Andrei Tokovinin}
\affil{Cerro Tololo Inter-American Observatory, Casilla 603, La Serena, Chile}
\email{atokovinin@ctio.noao.edu}

\author{David W. Latham} 
\affil{Harvard-Smithsonian Center for Astrophysics, 60
Garden Street, Cambridge, MA 02138, USA}
\email{dlatham@cfa.harvard.edu}

\author{Brian D. Mason}
\affil{U.S. Naval Observatory, 3450 Massachusetts Ave., Washington, DC, USA}
\email{bdm@usno.navy.mil}

\begin{abstract}
The young nearby  solar-type star HD 91962 is a  rare quadruple system
where three companions revolve  around the main component with periods
of 170.3  days, 8.84 years, and  205 years.  The two  outer orbits are
nearly co-planar, and all orbits have small eccentricities.  We refine
the  visual   orbit  of  the   outer  pair,  determine   the  combined
spectro-interferometric  orbit  of  the  middle 8.8-yr  pair  and  the
spectroscopic orbit of  the inner binary. The middle  and inner orbits
are likely  locked in  a 1:19  resonance, the ratio  of the  outer and
middle  periods  is  $\sim$23.   The  masses  of  all  components  are
estimated (inside-out:  1.14, 0.32,  0.64, 0.64 solar mass),  the dynamical
parallax is 27.4$\pm$0.6\,mas.  We speculate that this multiple system
originated from  collapse of an  isolated core and that  the secondary
components  migrated in  a dissipative  disk.  Other  multiple systems
with  similar features  (coplanarity, small  eccentricity,  and period
ratio around 20) are known.
\end{abstract}


\section{System description}
\label{sec:intro}

The  observed architectures  of  stellar systems  depend  both on  the
formation  processes  and subsequent  evolution.   In  many cases  the
orbits preserve  information about the formation  processes, and their
study helps us to  understand the physics of fragmentation, accretion,
and  early  evolution of  stars.   This  information  is gleaned  from
observations of young binaries, from statistics of binary and multiple
stars in different environments \citep{DK13}, and from unusual objects
that   reveal  the   history   of  their   formation  like   ``Rosetta
stones''. One such object is featured here.

\begin{figure}
\epsscale{1.0}
\plotone{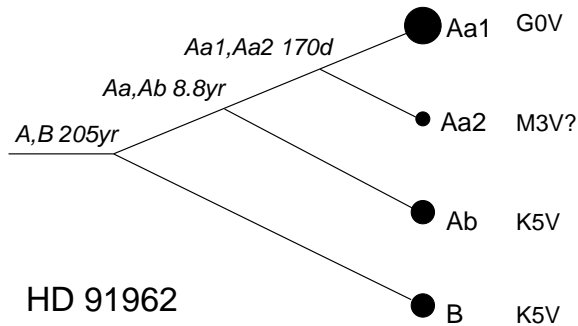}
\caption{Structure   of   the   hierarchical   quadruple   system   HD
  91962. Components are designated  by letters and numbers, subsystems
  are identified by their components joined by the comma.  Approximate
  spectral types  are assigned  to match the  estimated masses  of the
  stars.  All orbits have  small eccentricity and are possibly located
  in one plane.
\label{fig:str} 
}
\end{figure}

The  7th  magnitude G1V  star  studied  here  is known  as  HD~91962,
HIP~51966, WDS  J10370$-$0850, or ADS~7854; the  J2000 coordinates are
10:37:00.01,  $-$08:50:23.7.  It is  located at  a distance  of 36\,pc
from the  Sun.  HD~91962  is an X-ray  source RX~J1036.9$-$0850  and a
hierarchical quadruple system (Figure~\ref{fig:str}).

The outer  binary A,B (A~556) is known  since 1903 \citep{Aitken1904}.
Its orbit with $P=283$\,yr  by \citet{Pop1978} was recently revised by
\citet{TMH14}.  We  show below  that the period  is close  to 200\,yr.
The original  {\it Hipparcos}  parallax is 27.5$\pm$1.3\,mas,  the new
{\it    Hipparcos}    reduction    \citep{HIP2}    revised    it    to
25.1$\pm$1.2\,mas.  However,  the {\it  Hipparcos}  parallax could  be
biased by the orbital motion which was not taken into consideration in
its data reduction.

The  subsystem Aa,Ab  (TOK~44)  was discovered  by \citet{MH09}  with
adaptive optics in 2003.354,  at 0\farcs142, 56\fdg2, $\Delta K=1.25$ mag.
Independently, it  was resolved in  2009 by speckle  interferometry at
SOAR \citep{TMH10} and was measured several times since then. The pair
Aa,Ab  was  seen  in 2012.18  at  the  same  position as  in  2003.35,
completing one full revolution.   The orbital period is therefore well
constrained by the speckle measurements.

DL  independently determined the  spectroscopic orbit  of Aa,Ab
with  a period of  $3233 \pm  20$ days  (8.85\,yr) and  discovered the
inner  spectroscopic subsystem  Aa1,Aa2 with  $P=170.3$\,d.   This is
therefore a hierarchical quadruple  system with a 3-tier ``planetary''
architecture,  where all  components revolve  around the  most massive
central  star  Aa1.  The  two  inner  orbits  have small  and  similar
eccentricities  and their apsidal  angles are  also similar.   We show
below that  these orbits  are locked in  a 1:19 mutual  resonance. The
period of  the outer visual  orbit is about  20 times longer  than the
period of the middle orbit.

The  observational  material   is  presented  in
Section~\ref{sec:obs}. It is used for  calculation of the orbits in
Section~\ref{sec:orb} and  for the estimate of  component's masses and
distance   to  the  system   (Section~\ref{sec:mass}).  In   
Section~\ref{sec:disc}  we  discuss the  formation  mechanism of  such
hierarchies  and  give examples  of  other   3-tier  hierarchical
systems with known orbits. 

\section{Observations}
\label{sec:obs}

\subsection{Speckle interferometry}

\begin{figure}
\plotone{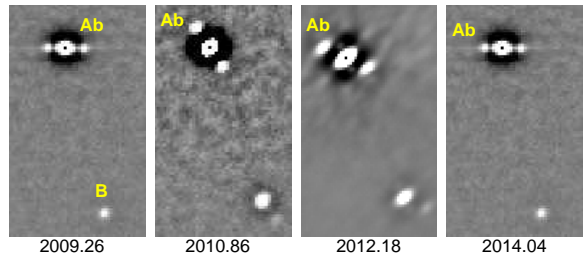} 
\caption{Fragments of  speckle auto-correlation functions  showing the
  resolved triple  system. The  scale and orientation  are approximate
  (North up, East  left). The two middle images  (2010.86 and 2012.18)
  are taken in  the $I_C$ band and contain  the faint peak corresponding
  to the cross-correlation  between B and Ab. In  the remaining images
  (in the $y$ filter) this peak is lost in the noise.
\label{fig:ACF} }
\end{figure}

All  resolved measurements of  the middle  subsystem Aa,Ab  except the
first one come from the  4.1-m SOAR telescope. The instrument and data
reduction are  described by \citet{TMH10}. The  observations were made
with the  534/22\,nm interference filter close to  the Str\"omgren $y$
band  and  the 788/132\,nm  filter  approximating  the Cousins  system
$I_C$.    Figure~\ref{fig:ACF}  presents   samples   of  the   speckle
auto-correlation  functions (ACFs)  of the  resolved triple  star. The
relative position  and brightness of  the components is  determined by
least-squares  fitting of the  power spectrum  to a  triple-star model
(not from the  ACF).  The orientation of the  inner pair is determined
without  the usual  $180^\circ$  ambiguity when  the correlation  peak
between Ab and B is detectable  (it is barely seen in the $y$ filter).
Three  observations  made  in   2014  are  still  unpublished.   After
submission  of the manuscripot,  two more  measurenments made  in 2015
were added, slightly reducing the errors of the middle orbit.  We also
use one unpublished  measure of A,B made in 2001  by BM (the subsystem
was not resolved).

\subsection{Spectroscopy}

\begin{deluxetable}{ r rrr l }          
\tabletypesize{\scriptsize}     
\tablecaption{Parameters of  the star A\rm{a}1
\label{tab:Teff}  }
\tablewidth{0pt}      
\tablehead{  
\colhead{$T_e$} &
\colhead{$\log g$} &
\colhead{[m/H]} &
\colhead{$V \sin i$} &
\colhead{Reference} \\
(K) & 
(m~s$^{-2}$) & 
(Sun) & 
km~s$^{-1}$  & 
}
\startdata
5827 & 4.49 & $-$0.15 & 9.27    & This work \\
$\pm 17$ & $\pm 0.01$ & $\pm 0.01$ & $\pm 0.26$ & \\
5818 & 4.85 & \ldots & 6.10 & \citet{Schroeder} \\
5675 & 4.12 & $-$0.21 & \ldots &  \citet{Casagrande2011} 
\enddata
\end{deluxetable}

The spectrum of HD 91962 was  monitored for 23 years starting in 1991.
Eighty-four   observations  were   obtained  with   the   CfA  Digital
Speedometers \citep{Latham1985,Latham1992},  initially using the 1.5-m
Wyeth Reflector at  the Oak Ridge Observatory in  the town of Harvard,
Massachusetts, and  subsequently with the  1.5-m Tillinghast Reflector
at  the Whipple Observatory  on Mount  Hopkins, Arizona.   Starting in
2009  the  new fiber-fed  Tillinghast  Reflector Echelle  Spectrograph
\citep[TRES;][]{TRES}   was   used  to   obtain   an  additional   ten
observations.   The  spectral  resolution  was 44,000  for  all  three
spectrographs,  but  the   typical  signal-to-noise  ratio  (SNR)  per
resolution element  of 100 for the  TRES observations was  a few times
higher than for the CfA Digital Speedometer observations.

The total light of HD 91962  is dominated by the component Aa1, and at
visible wavelengths  the spectrum appears  single-lined.  Therefore we
followed our standard  procedure of using one-dimensional correlations
of each observed spectrum against  a synthetic template drawn from our
library of  calculated spectra.  The  radial velocity (RV)  zero point
for  each spectrograph  was monitored  using observations  of standard
stars, of daytime  sky, and of minor planets,  and the velocities were
all adjusted to the native  system of the CfA Digital Speedometers. To
get onto the  absolute velocity system defined by  our observations of
minor  planets,  about  0.14   km~s$^{-1}$  should  be  added  to  the
velocities  reported in Table~3.   These velocities  are all  based on
correlations  of just  a single  echelle order  centered on  the  Mg b
triplet near 519\,nm, with a  wavelength window of 4.5\,nm for the CfA
Digital Speedometers and 10.0\,nm for TRES.

The  ten TRES observations  were analyzed  with the  Stellar Parameter
Classification  tool  \citep[SPC;][]{Bucchave2014}.   The results  are
given in  Table~\ref{tab:Teff}.  As  they are mutually  consistent, we
list  the  average  values  of  stellar parameters  and  their  errors
estimated from the scatter  of ten measurements.  The parameters found
in the literature are given  for comparison. The star Aa1 dominates in
the  visible light,  with all  other components  together contributing
only 14\%. Therefore their influence on the derived stellar parameters
of Aa1 is minor.

\subsection{Photometry}

\begin{deluxetable}{c  ccc  ccc }
\tablecaption{Photometry of components
\label{tab:ptm1} }
\tablewidth{0pt}                                                                           
\tablehead{  
\colhead{Band} &
\colhead{A+B} &
\colhead{Ab$-$Aa} &
\colhead{B$-$A} &
\colhead{Aa} &
\colhead{Ab} &
\colhead{B} 
}
\startdata
$V$ (mag)   & 7.03 &  2.86$\pm$0.10  & 2.67$\pm$0.10  & 7.19 & 10.05 & 9.80\\
$I_C$ (mag) & 6.34 &  2.29$\pm$0.10  & 2.09$\pm$0.05  & 6.62 & 8.91 & 8.55 \\
$K_s$ (mag) & 5.39 &  1.25$\pm$0.11  & 1.37$\pm$0.06  & 5.96 & 7.21 & 7.03 \\
Mass (${\cal M}_\odot$)  & 2.74 & \ldots     & \ldots   & 1.46 & 0.64 & 0.64 
\enddata
\end{deluxetable}

Table~\ref{tab:ptm1} gathers the  photometric information.  Its column
(2) gives the  combined magnitudes of AB, with  $V$ from SIMBAD, $I_C$
calculated from the $V-I$ color  given in \citep{HIP2}, and $K_s$ from
2MASS \citep{2MASS}.  The magnitude  differences between Ab and Aa and
between B and A are given in  the columns (3) and (4).  They are based
on speckle  interferometry at SOAR, assuming $\Delta  V \approx \Delta
y$, and on photometry from  \citep{MH09}: $\Delta K_{\rm Aa,Ab} = 1.25
\pm  0.11$ mag  and $\Delta  K_{\rm  AB} =  1.37 \pm  0.06$ mag.   The
scatter of the relative speckle  photometry is about 0.1\,mag, and the
magnitude difference between Aa and B is slightly over-estimated owing
to the  anisoplanatism.  This bias  is overcome by using  the relative
photometry of the wide pair  on the long-exposure images produced from
the speckle  data cubes: $\Delta y_{\rm  AB} = 2.67 \pm  0.10$ mag and
$\Delta I_{\rm AB}  = 2.09 \pm 0.05$ mag, in  good agreement with {\it
  Hipparcos}  and {\it  Tycho} \citep{FM2000}:  $\Delta Hp_{\rm  AB} =
2.69$, $\Delta V_{\rm AB} = 2.48$  and $\Delta B_{\rm AB} = 3.30$ mag.
Note  also the  magnitude difference  $\Delta V_{\rm  AB} =  2.52$ mag
measured by \citet{Horch2001}.

The  last  three   columns  of  Table~\ref{tab:ptm1}  list  individual
magnitudes calculated  from the combined  and differential photometry.
The component Aa  is treated as a  single star, while it is  in fact a
170-d  binary.  Masses  given  in  the  last  line  are  discussed  in
Section~\ref{sec:mass}.

\section{Orbits}
\label{sec:orb}

\begin{deluxetable}{l l  cc c  }          
\tablecaption{Orbital elements of HD 91962.
\label{tab:orb} }
\tablewidth{0pt}   
\tablehead{  
\colhead{Element} &
\colhead{A,B} &
\colhead{Aa,Ab} &
\colhead{Aa1,Aa2} 
}
\startdata 
$P$ (yr)              & 205  (fixed)     & 8.84638 $\pm$ 0.025 & 0.46628 $\pm$ 0.00005    \\ 
$P$ (d)               & 74875  (fixed)     & 3230.8 $\pm$ 9       & 170.304 $\pm$ 0.013  \\ 
$T$ (yr)              & 2048.6 $\pm$ 2.5  &2009.816 $\pm$ 0.072 &  2003.292 $\pm$ 0.005  \\    
$T$ (JD + 2,400,0000) & 69290 $\pm$ 873   &55129   $\pm$ 26    &  52746.80 $\pm$ 1.48 \\    
$e$                  & 0.301 $\pm$ 0.016 &0.125 $\pm$ 0.010    &  0.135 $\pm$ 0.008 \\  
$a$ (arcsec)          & 1.334 $\pm$ 0.016 &0.1501 $\pm$ 0.0021    &  (0.0184) \\   
$\Omega_A$ (deg)    & 63.2 $\pm$ 2.0    &50.4 $\pm$ 1.0       &  \ldots \\  
$\omega_A$ (deg)    & 219.7 $\pm$ 1.3  &263.9 $\pm$ 2.6      &  297.6 $\pm$  3.2 \\  
$i$   (deg)         & 54.2 $\pm$ 0.8   &56.6 $\pm$ 0.9      & (57) \\  
$K_1$ (km~s$^{-1}$)            & (0.21)         & 4.71$\pm$ 0.07     & 8.08 $\pm$   0.07 \\
$V_0$ (km~s$^{-1}$)            & \ldots             & \ldots                 & 21.16 $\pm$ 0.04   
\enddata
\end{deluxetable}

\begin{deluxetable}{ cc cr  cc cr  }          
\tabletypesize{\scriptsize}     
\tablecaption{Radial velocities and residuals 
\label{tab:RV}  }
\tablewidth{0pt}                                                                           
\tablehead{  
\colhead{JD} &
\colhead{RV} &
\colhead{Err} &
\colhead{O$-$C} &
\colhead{JD} &
\colhead{RV} &
\colhead{Err} &
\colhead{O$-$C} \\
\colhead{+2,400,000} & 
\colhead{(km~s$^{-1}$)} & 
\colhead{(km~s$^{-1}$)} & 
\colhead{(km~s$^{-1}$)} & 
\colhead{+2,400,000} & 
\colhead{(km~s$^{-1}$)} & 
\colhead{(km~s$^{-1}$)} & 
\colhead{(km~s$^{-1}$)}       
}
\startdata
 48290.835  &    12.11 &   0.44  &    0.35 &   53838.722  &    19.32 &   0.21 &  --0.07 \\
 49051.810  &    28.68 &   0.43  &  --0.35 &   53866.741  &    13.20 &   0.19 &  --0.32 \\
 49165.488  &    27.95 &   0.39  &    0.22 &   53871.656  &    12.16 &   0.25 &  --0.63 \\
 49384.843  &    32.55 &   0.32  &    0.64 &   54071.034  &    10.39 &   0.23 &  --0.39 \\
 49391.711  &    31.52 &   0.39  &    1.09 &   54077.023  &    11.65 &   0.20 &    0.05 \\
 49479.349  &    19.67 &   0.41  &  --0.03 &   54100.952  &    18.56 &   0.25 &  --0.22 \\
 49480.348  &    19.75 &   0.42  &  --0.14 &   54107.933  &    21.17 &   0.22 &  --0.21 \\
 49483.360  &    20.42 &   0.37  &  --0.12 &   54127.920  &    25.35 &   0.36 &  --0.56 \\
 49708.698  &    34.25 &   0.49  &    0.68 &   54135.897  &    26.03 &   0.21 &    0.07 \\
 49748.613  &    25.63 &   0.41  &  --0.05 &   54158.887  &    22.25 &   0.18 &  --0.03 \\
 49750.574  &    25.13 &   0.37  &  --0.07 &   54166.821  &    20.25 &   0.25 &  --0.15 \\
 49775.520  &    20.33 &   0.55  &    0.48 &   54187.787  &    15.59 &   0.28 &    0.22 \\
 49807.431  &    16.69 &   0.41  &  --0.48 &   54192.724  &    13.92 &   0.42 &  --0.36 \\
 49813.412  &    17.03 &   0.45  &  --0.56 &   54196.739  &    12.68 &   0.27 &  --0.77 \\
 50070.693  &    27.17 &   0.41  &  --1.03 &   54221.689  &    10.57 &   0.24 &    0.76 \\
 50128.539  &    16.18 &   0.43  &    0.21 &   54251.721  &    11.73 &   0.28 &    0.08 \\
 50138.507  &    15.10 &   0.49  &  --0.04 &   54277.656  &    20.74 &   0.29 &    0.35 \\
 50161.454  &    16.20 &   0.36  &  --0.37 &   54457.972  &    22.99 &   0.21 &  --0.12 \\
 50163.476  &    17.20 &   0.38  &    0.23 &   54461.968  &    25.02 &   0.23 &    1.05 \\
 50170.394  &    18.79 &   0.49  &    0.05 &   54481.934  &    24.54 &   0.23 &    0.08 \\
 50183.390  &    22.96 &   0.38  &  --0.29 &   54486.927  &    24.67 &   0.22 &    0.88 \\
 50185.384  &    23.89 &   0.43  &  --0.12 &   54516.861  &    17.48 &   0.23 &    0.26 \\
 50189.385  &    25.06 &   0.45  &  --0.44 &   54521.883  &    16.28 &   0.35 &    0.24 \\
 50458.658  &    15.34 &   0.37  &    0.27 &   54543.743  &    11.82 &   0.29 &    0.30 \\
 50460.619  &    15.59 &   0.43  &    0.84 &   54549.829  &    10.93 &   0.25 &    0.37 \\
 50484.559  &    12.56 &   0.44  &  --0.02 &   54576.693  &     8.38 &   0.22 &  --0.62 \\
 50514.480  &    17.80 &   0.42  &    0.41 &   54601.702  &    14.80 &   0.32 &    0.91 \\
 50516.461  &    18.47 &   0.43  &    0.42 &   54605.617  &    15.23 &   0.35 &  --0.04 \\
 52997.868  &    21.85 &   0.28  &  --0.67 &   54808.035  &    25.21 &   0.22 &  --0.53 \\
 53037.771  &    17.90 &   0.32  &    0.83 &   54839.907  &    22.61 &   0.29 &  --0.02 \\
 53054.679  &    18.59 &   0.27  &    0.29 &   54846.883  &    20.91 &   0.22 &  --0.16 \\
 53087.635  &    28.91 &   0.35  &    0.21 &   54868.897  &    16.35 &   0.22 &    0.28 \\
 53106.657  &    33.38 &   0.34  &    0.65 &   54878.903  &    13.67 &   0.26 &  --0.41 \\
 53353.876  &    17.92 &   0.28  &    0.46 &   54898.830  &    10.77 &   0.28 &  --0.48 \\
 53388.852  &    15.65 &   0.36  &    0.29 &   54924.741  &    11.15 &   0.30 &  --0.46 \\
 53403.862  &    18.33 &   0.22  &    0.29 &   54957.690  &    22.20 &   0.30 &    0.38 \\
 53418.705  &    23.45 &   0.23  &    0.43 &   54962.655  &    23.28 &   0.31 &  --0.35 \\
 53433.695  &    27.45 &   0.27  &  --0.78 &   55169.033  &    28.08 &   0.10 &  --0.25 \\
 53448.694  &    30.38 &   0.26  &  --0.21 &   55172.031  &    27.66 &   0.10 &  --0.18 \\
 53465.621  &    28.12 &   0.46  &  --1.14 &   56743.750  &    19.17 &   0.10 &  --0.31 \\
 53485.588  &    24.59 &   0.26  &  --0.31 &   56790.638  &    13.60 &   0.10 &  --0.23 \\
 53510.706  &    19.19 &   0.19  &    0.29 &   56804.634  &    15.39 &   0.10 &  --0.14 \\
 53541.647  &    14.13 &   0.26  &    0.05 &   56811.653  &    17.16 &   0.10 &  --0.09 \\
 53721.004  &    12.61 &   0.52  &    0.07 &   56817.658  &    19.05 &   0.10 &  --0.10 \\
 53746.976  &    16.04 &   0.33  &  --0.27 &   56820.637  &    19.99 &   0.10 &  --0.21 \\
 53776.914  &    26.13 &   0.45  &  --0.34 &   56825.656  &    22.02 &   0.10 &  --0.06 \\
 53836.847  &    19.84 &   0.21  &    0.00 &   56827.656  &    22.60 &   0.10 &  --0.23 \\
\enddata
\end{deluxetable}

\begin{deluxetable}{ rrr c rr }          
\tabletypesize{\scriptsize}     
\tablecaption{Measurements and residuals of A\rm{a},A\rm{b}
\label{tab:Aa}  }
\tablewidth{0pt}                                                                           
\tablehead{  
\colhead{Date} &
\colhead{$\theta$} &
\colhead{$\rho$} &
\colhead{$\sigma_\rho$} &
\colhead{(O$-$C)$_\theta$} &
\colhead{(O$-$C)$_\rho$} \\
  & 
\colhead{(deg)} & 
\colhead{(arcsec)} & 
\colhead{(arcsec)} &    
\colhead{(deg)} & 
\colhead{(arcsec)}
}
\startdata
  2003.354 &   56.2 &  0.142 &  0.005 & $-$2.5 & $-$0.009   \\
  2009.263 &  268.8 &  0.097 &  0.005 &    0.1 &    0.000   \\
  2010.969 &   29.0 &  0.122 &  0.005 &    1.6 &    0.003   \\
  2010.969 &   30.0 &  0.125 &  0.005 &    2.5 &    0.006   \\
  2012.184 &   58.0 &  0.151 &  0.005 & $-$0.3 &    0.000   \\
  2012.184 &   58.1 &  0.151 &  0.005 & $-$0.2 &    0.000   \\
  2013.132 &   81.9 &  0.124 &  0.005 &    0.1 & $-$0.005   \\
  2013.132 &   83.8 &  0.135 &  0.105 &    1.9 &    0.006   \\
  2014.043 &  117.7 &  0.097 &  0.002 & $-$1.1 &    0.000   \\
  2014.186 &  126.6 &  0.097 &  0.002 &    0.1 &    0.003   \\
  2014.300 &  133.0 &  0.097 &  0.002 &    0.0 &    0.004   \\
  2015.029 &  172.1 &  0.098 &  0.002 & $-$0.9 & $-$0.005   \\
  2015.169 &  181.0 &  0.108 &  0.002 &    1.6 &    0.001   
\enddata
\end{deluxetable}

\begin{deluxetable}{ rrr c rr }          
\tabletypesize{\scriptsize}     
\tablecaption{Measurements and residuals of A,B
\label{tab:AB}  }
\tablewidth{0pt}                                                                           
\tablehead{  
\colhead{Date} &
\colhead{$\theta$} &
\colhead{$\rho$} &
\colhead{$\sigma_\rho$} &
\colhead{(O$-$C)$_\theta$} &
\colhead{(O$-$C)$_\rho$} \\
  & (deg) & (arcsec) & (arcsec) &    (deg) & (arcsec)
}
\startdata
  1903.040 &   54.0 &  1.340 &  0.500 &   $-$5.3 & $-$0.185   \\
  1909.320 &   65.2 &  1.800 &  0.100 &    1.4 &  0.214   \\
  1909.320 &   65.0 &  1.640 &  0.100 &    1.2 &  0.054   \\
  1911.100 &   60.4 &  1.600 &  0.100 &   $-$4.6 &  0.001   \\
  1912.046 &   66.2 &  1.650 &  0.100 &    0.6 &  0.045   \\
  1915.210 &   69.6 &  1.480 &  0.100 &    1.8 & $-$0.140   \\
  1916.210 &   70.0 &  1.700 &  0.100 &    1.6 &  0.076   \\
  1921.360 &   73.1 &  1.610 &  0.100 &    1.3 & $-$0.023   \\
  1925.340 &   76.8 &  1.680 &  0.100 &    2.4 &  0.050   \\
  1926.260 &   77.0 &  1.550 &  0.100 &    2.0 & $-$0.078   \\
  1928.160 &   81.3 &  1.520 &  0.100 &    5.0 & $-$0.102   \\
  1929.510 &   74.2 &  1.540 &  0.100 &   $-$3.0 & $-$0.077   \\
  1933.280 &   78.0 &  1.560 &  0.100 &   $-$1.7 & $-$0.038   \\
  1933.960 &   76.5 &  1.330 &  0.500 &   $-$3.7 & $-$0.264   \\
  1936.250 &   81.4 &  1.350 &  0.100 &   $-$0.4 & $-$0.228   \\
  1939.300 &   81.5 &  1.520 &  0.100 &   $-$2.4 & $-$0.034   \\
  1944.010 &   86.4 &  1.450 &  0.100 &   $-$1.0 & $-$0.059   \\
  1944.290 &   92.4 &  1.400 &  0.100 &    4.7 & $-$0.106   \\
  1947.930 &   89.0 &  1.280 &  0.100 &   $-$1.5 & $-$0.184   \\
  1956.340 &   97.6 &  1.280 &  0.100 &   $-$0.3 & $-$0.074   \\
  1957.760 &  106.2 &  1.520 &  0.100 &    6.9 &  0.187  \\
  1958.000 &   97.0 &  1.420 &  0.100 &   $-$2.5 &  0.091   \\
  1958.040 &  103.1 &  1.380 &  0.100 &    3.5 &  0.051   \\
  1959.160 &  110.8 &  1.600 &  1.100 &   10.1 &  0.285   \\
  1959.300 &   97.7 &  1.140 &  0.100 &   $-$3.1 & $-$0.170   \\
  1962.500 &  110.3 &  1.180 &  0.100 &    6.1 & $-$0.082   \\
  1966.380 &  108.4 &  1.200 &  0.100 &   $-$0.3 &  $-$0.002   \\
  1972.112 &  116.5 &  1.140 &  0.100 &    0.4 &  0.026   \\
  1972.120 &  113.1 &  0.920 &  0.100 &   $-$3.0 & $-$0.194   \\
  1977.280 &  121.3 &  1.160 &  0.100 &   $-$2.6 &  0.123   \\
  1982.260 &  127.9 &  1.010 &  0.100 &   $-$4.1 &  0.039   \\
  1991.250 &  151.7 &  0.886 &  0.010 &    0.7 &  0.001   \\
  1996.350 &  141.7 &  0.723 &  1.100 &  $-$21.0 & $-$0.138   \\
  1997.123 &  165.6 &  0.833 &  0.020 &    1.1 & $-$0.027   \\
  1997.300 &  141.4 &  0.753 &  1.100 &  $-$23.5 & $-$0.106   \\
  2001.077 &  172.4 &  0.842 &  0.005 &   $-$1.5 & $-$0.017   \\
  2002.167 &  177.3 &  0.873 &  0.005 &    0.9 &  0.012   \\
  2009.263 &  192.6 &  0.906 &  0.005 &    0.0 &  0.013   \\
  2010.969 &  196.0 &  0.899 &  0.005 &   $-$0.3 & $-$0.006   \\
  2010.969 &  196.0 &  0.904 &  0.005 &   $-$0.3 & $-$0.001   \\
  2012.184 &  198.7 &  0.909 &  0.005 &   $-$0.2 & $-$0.005   \\
  2013.132 &  200.7 &  0.924 &  0.005 &  $-$0.0 &  0.004   \\
  2013.132 &  200.9 &  0.926 &  0.005 &    0.1 &  0.006   \\
  2014.043 &  203.1 &  0.923 &  0.005 &    0.4 & $-$0.005  \\
  2014.186 &  202.9 &  0.925 &  0.005 & $-$0.1 & $-$0.003   \\
  2014.300 &  203.5 &  0.933 &  0.005 &    0.3 &  0.004   \\
  2015.029 &  204.7 &  0.939 &  0.005 &  $-$0.0 &  0.004 \\
  2015.169 &  205.4 &  0.937 &  0.005 &    0.4  & 0.001  
\enddata
\end{deluxetable}

\begin{figure*}
\epsscale{2.0}
\plotone{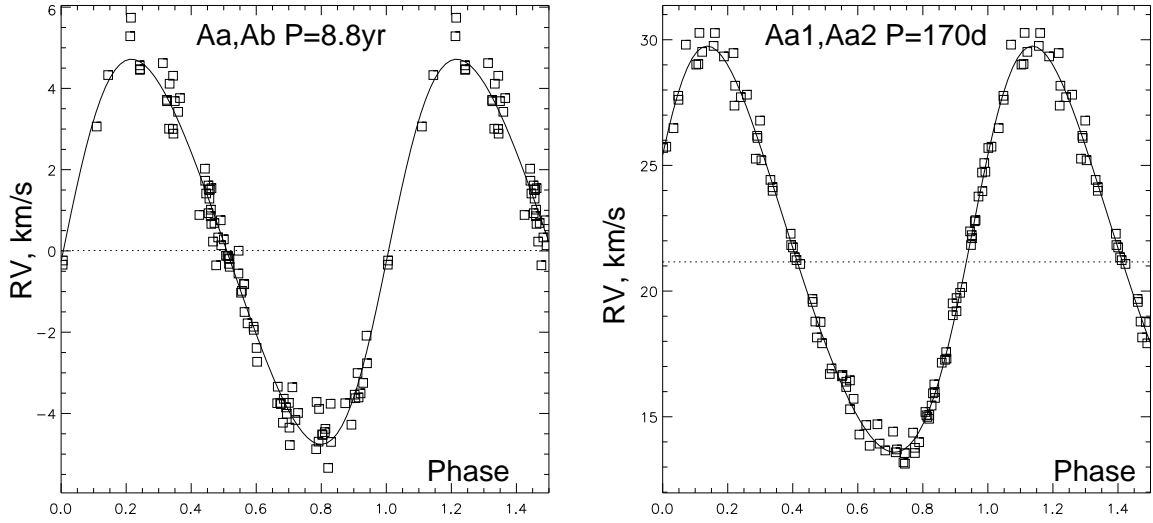}
\caption{Spectroscopic orbits  of the middle (left) and the inner (right)
  subsystems of HD~91962.\label{fig:sborb} 
}
\end{figure*}

\begin{figure*}
\epsscale{2.0}
\plotone{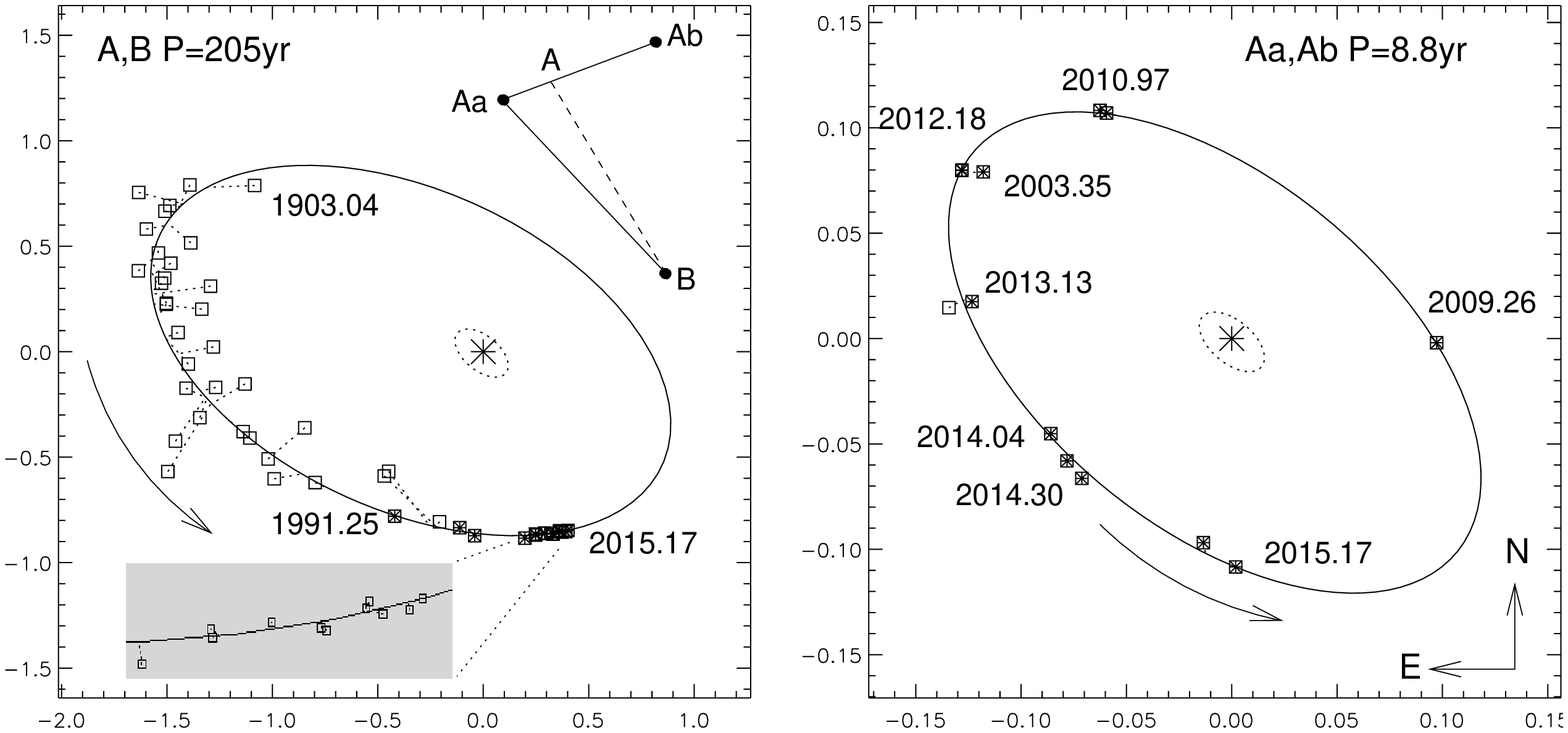}
\caption{Visual orbits of  the outer system A,B (left)  and the middle
  system Aa,Ab (right).  The scale is in arcseconds.  The small dotted
  ellipses  shows inner  orbits to  scale  (Aa1,Aa2 is  assumed to  be
  co-planar with Aa,Ab). The scheme  in the left panel illustrates the
  correction  of the resolved  measurements of  Aa,B using  the vector
  formula $\vec{Aa,B} = \vec{Aa,A} + \vec{A,B}$.
\label{fig:vborb} 
}
\end{figure*}

Table~\ref{tab:orb} lists  the orbital  elements and their  errors for
all three orbits:  outer, middle, and inner. In  purely visual orbits,
the ascending node  is not known, meaning that  both elements $\Omega$
and $\omega$ can be changed  by $180^\circ$. Radial velocities help to
select the correct node.   Here the longitude of periastron $\omega_A$
corresponds to  the primary component,  and the position angle  of the
node $\Omega_A$  is chosen in such  way as to represent  the motion of
the secondary on its visual orbit. Estimated or assumed quantities are
given in brackets for reference. The orbital elements and their errors
are  determined by  the unconstrained  least-squares fit  to  the data
(RVs,  positional  measurements,   or  both)  with  weights  inversely
proportional  to   the  squares   of  the  measurement   errors.   The
center-of-mass velocity is  common to all three orbits,  but we do not
know  yet the spectroscopic  elements of  A,B and  arbitrarily ascribe
$V_0$ to the inner pair Aa1,Aa2 meanwhile.

The  RVs used in  the orbit  calculation and  common residuals  to the
orbits of Aa1,Aa2 and  Aa,Ab are presented in Table~\ref{tab:RV}.  The
speckle measurements of Aa,Ab and residuals to its orbit are listed in
Table~\ref{tab:Aa}. The first measurement is made by \citet{MH09}, all
remaining measurements  are made at  SOAR.

The  orbit   of  the  middle  pair  Aa,Ab   was  initially  determined
independently from  both speckle interferometry and RVs.   We used the
initial spectroscopic  orbits of Aa,Ab  and Aa1,Aa2 fitted  jointly to
the RV  data alone as a  first approximation to the  combined orbit of
Aa,Ab (the 170-d  orbit was then subtracted from  the RVs).  With this
combined  solution, another iteration  on the  inner system  was made,
giving essentially the same elements.  The weighted rms RV residual to
both orbits is 0.42\,km~s$^{-1}$.  Figure~\ref{fig:sborb} shows the RV
curves, while Figure~\ref{fig:vborb} shows  the visual orbits of Aa,Ab
and A,B.

The final  elements of  Aa,Ab and their  errors are determined  by the
least-squares  fit to  both  RVs and  resolved  measures. The  speckle
errors are  assumed to be  5\,mas prior to  2014 (except for  one less
precise measure).   and 2\,mas for 2015--2015.  The  rms residuals are
1\fdg2 and  3.4\,mas in $\theta$  and $\rho$, respectively. We  had to
adopt realistic  speckle errors (hence  weights) to reach  the correct
balance between  positional measurements  and RVs.  The  formal errors
delivered by the speckle  data processing are smaller, typically under
1\,mas.

We updated the  visual orbit of the outer system A,B  = A~556 (see the
elements  in   Table~\ref{tab:orb},  obseravtions  and   residuals  in
Table~\ref{tab:AB}).  The speckle measures of Aa,B (all data from SOAR
where the  triple is  resolved) are translated  into the  positions of
A,B, where A  refers to the center of mass of  Aa,Ab.  The position of
the inner pair Aa,Ab was calculated from its kown orbit and the vector
directed from Aa to Ab was added to the vector Aa,B with a coefficient
$\alpha = -q/(1+q) = -0.305$ ($q  = 0.44$ is the mass ratio in Aa,Ab).
The unresolved accurate  measurements (1991--2001) were corrected with
$\alpha = -(q -r)/[(1+q)(1+r)]  = -0.237$, considering that they refer
to the  photo-center of Aa,Ab  and not  to A or  Aa ($r =0.07$  is the
light ratio Ab/Aa).  This correction substantially reduces the scatter
of  accurate  speckle measurements  (see  the  scheme  and the  zoomed
portion  of the  orbit in  Figure~\ref{fig:vborb}, left).   The visual
micrometer measurements  are left uncorrected, as the  effect of Aa,Ab
is $<36$\,mas.  They are assigned  errors of 0\farcs1, except the four
highly deviant  micrometer measurements that were  given larger errors
to cancel their influence on the orbit.

It turns  out that the data do  not yet constrain all  elements of the
outer pair A,B.  Equally good solutions can be  obtained by fixing the
period  at different  values within  a certain  range;  longer periods
correspond  to a  smaller  mass  sum.  The  orbit  given here  assumes
$P=205$\,yr  and gives the  mass sum  of 2.68  ${\cal M}_\odot$  for a
parallax  of 27.6\,mas.  The  unconstrained  fit gives  $P  = 240  \pm
35$ years.

Using the masses  estimated below, we calculate that  the RV amplitude
in the A,B orbit is  $K_1 = 0.2$\,km~s$^{-1}$.  The ephemeris predicts
that the RV(A) should change by $-0.18$\,km~s$^{-1}$ during the period
1991--2014 covered by the observations.  We fitted the RV residuals by
a  linear   function  and  found   the  coefficient  of   $-0.007  \pm
0.008$\,km~s$^{-1}$~yr$^{-1}$,     or     a     total    change     of
$-0.17$\,km~s$^{-1}$ during the 23-yr  period of RV observations.  The
concidence of those numbers is  accidental, given that the RV trend is
not formally significant.  However, the  {\it sign} of the emerging RV
trend tells  us that the node of  the orbit of A,B  is probably chosen
correctly.   Therefore,  the  orbits  of  A,B  and  Aa,Ab  are  nearly
co-planar (the angle between the angular momenta is $\phi = 10.8^\circ
\pm  1.9^\circ$).   If the  node  of the  outer  orbit  is changed  by
$180^\circ$, then  $\phi = 110^\circ$.  In such  case, the Kozai-Lidov
cycles would  have made  the middle orbit  highly eccentric  and would
have destroyed the architecture of this multiple system.

The calculated semi-major axis of Aa1,Aa2 is 18.4\,mas. The ``wobble''
of Aa due  to the inner subsystem should have  an amplitude of 4\,mas.
The  orientation of  the inner  orbit  Aa1,Aa2 can  be established  by
frequent and precise speckle  measurements of Aa,Ab.  The measurements
of Aa,Ab made in 2014--2015 seem to deviate from the middle orbit in a
systematic way, but we could not yet use the residuals for determining
the elements $\Omega$  and $i$ of the inner orbit.  The reason is that
the  number of measurements  of Aa,Ab  is still  modest, leading  to a
cross-talk between the  elements of the middle and  inner orbits.  The
co-planarity of those orbits  thus remains hypothetical.  However, the
moderate  eccentricity  of the  inner  orbit  implies  the absence  of
Kozai-Lidov cycles, hence mutual inclination $\phi < 39^\circ$.

An attempt was made to measure the RVs of the faint components Ab and B
using  two-dimensional   correlation,  TODCOR  \citep{TODCOR}.   Three
well-exposed spectra from TRES  (JD 2455169 to 2456743) were processed
using synthetic  templates with  effective temperatures $T_e$  of 5750
and 4500\,K.  A second maximum was seen at velocities of 21.74, 21.44,
and 25.07 km~s$^{-1}$.   All three dates are close to  the node of the
middle orbit  Aa,Ab, so the  measured velocities of the  secondary, if
real, correspond to a blend between Ab and B.  The measurements should
be repeated in a couple of years, at a different  phase of the middle orbit.

\section{Masses and distance}
\label{sec:mass}

\begin{figure}
\epsscale{1.0}
\plotone{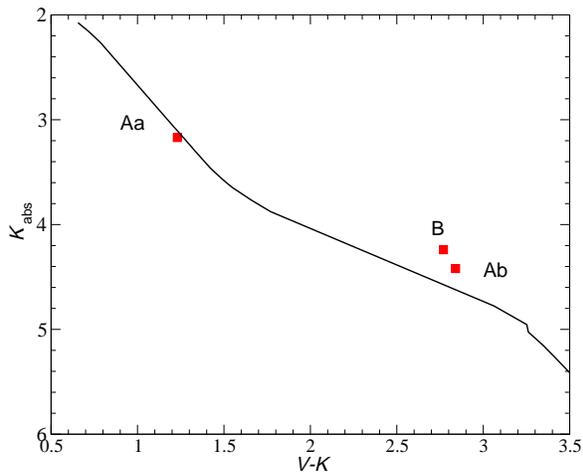}
\caption{Components Aa, Ab,  and B on the $(K_s, V -  K_s)$ CMD with a
  parallax of 27.6\,mas. The  line shows the 1-Gyr Dartmouth isochrone
  with solar metallicity \citep{Dotter2008}.
\label{fig:cmd} 
}
\end{figure}

The masses and distance  were determined iteratively, using photometry
and orbital elements. Several assumptions  are made: (i) the stars are
not  evolved and  follow standard  mass-luminosity relations  for main
sequence stars;  (ii) the  component  Aa2 contributes  little  light in  all
bands,  and  (iii) the  inner  orbit  Aa1,Aa2  has an  inclination  of
$57^\circ$.

Using  the initial  estimates  of masses  from  the absolute  $V$-band
magnitudes, we determine the  dynamical parallax from the middle orbit
Aa,Ab, then refine the masses and other parameters.  The numbers given
below were obtained in the last iteration.

The  mass of  Aa1 is  1.14 ${\cal  M}_\odot$, as  determined  from the
isochrone in Figure~\ref{fig:cmd}.  Then  the orbit of Aa1,Aa2 and the
assumption (iii) lead to the  mass of Aa2 of 0.32${\cal M}_\odot$. The
mass of Aa is therefore 1.46${\cal M}_\odot$.  The orbit of Aa,Ab with
known inclination  leads to  the mass of  0.64 $M_\odot$ for  Ab.  The
mass of  A=(Aa+Ab) is therefore  established at 2.10  ${\cal M}_\odot$
and its semi-major axis is  5.48\,AU.  The semi-major axis of Aa,Ab is
measured at  $150 \pm 2$\,mas and  leads to the  dynamical parallax of
$27.4  \pm  0.6$\,mas, i.e.  a  2\%  accuracy  on the  distance.   The
distance is proportional to the mass  sum to 1/3 power, so revision of
the  mass  estimates  would  have  a minor  effect  on  the  dynamical
parallax.

Using the distance and photometry, we place the components Aa, Ab, and
B on the color-magnitude diagram (Figure~\ref{fig:cmd}).  The absolute
magnitude  of B  corresponds to  a main-sequence  star  of 0.74~${\cal
  M}_\odot$.  However,  the  components Ab  and  B have  similar
luminosity, so we adopt the mass of 0.64~${\cal M}_\odot$ for B, close
to  the  measured value  for  Ab.  The mass  sum  of  A+B is  therefore
2.74~${\cal M}_\odot$.  The orbit of A,B  with a fixed period of
205\,yr and the parallax of 27.4\,mas corresponds to such a mass sum.

\section{Discussion}
\label{sec:disc}

The four known components of  the quadruple system HD~91962 are normal
dwarfs that  match the  standard mass-luminosity relation.   We cannot
exclude additional  low-mass satellites revolving  around Ab or  B, as
their RVs were  not measured directly, while the  constraints from the
photometry and orbits are not tight enough.

The system  appears to be young.  \citet{White07}  measured the lithum
6707\AA~ line strength of 73\,m\AA, axial rotation of 17\,km~s$^{-1}$,
and  detected  chromospheric  emission  in a  spectrum  with  a
resolution of 16,000.   \citet{Schroeder} confirmed the chromospheric
emission, explaining the X-ray flux.   The star does not belong to any
known kinematic group.  No excess far-infrared emission was found with
{\em Spitzer} \citep{Carpenter2009}.  Indeed,  a debris disk would not
survive inside this  multiple system.  If the dust  exists outside the
orbit of A,B it would be too cold to be detectable.

All three orbits  have low eccentricity.  The orbits  of A,B and Aa,Ab
have mutual inclination of $11^\circ$.  Small eccentricities imply the
absence  of  the  Kozai-Lidov  cycles,  hence  moderate  mutual  orbit
inclinations at all hierarchical  levels.  Moreover, the longitudes of
periastron $\omega$ in all three orbits are also similar, showing that
the lines of apsides have  similar orientation.  One can't help noting
the  similarities of  the orbits  while comparing  the left  and right
parts of Figures~\ref{fig:sborb} and \ref{fig:vborb}. 

The period  ratio between the middle  and inner systems  is $18.97 \pm
0.06$.  It appears that these orbits are in a weak 1:19 resonance. The
period ratio  of the outer and middle  systems is about 23  (it is not
accurate enough  to check for  a resonance).  The quadruple  system is
dynamically stable  and is organized  in a regular way,  remiscient of
the Solar  system (Figure~\ref{fig:str}).  Similar  eccentricities and
apsidal angles, as well as  the resonance, suggest that the companions
interacted with each other  during their formation and early dynamical
evolution.

This  quadruple  system  could   originate  in  an  isolated  rotating
core. Rotation prevented immediate  collapse. The gas formed a massive
and  unstable  disk which  fragmented  into  a companion.   Continuing
accretion  onto the  companion increased  its mass  and  caused inward
migration, while  dissipative gas friction maintained  the low orbital
eccentricity.  The first companions could have merged with the central
body  Aa1, while  other companions  were formed  on the  periphery and
migrated inwards.  The process was  stopped when the gas reservoir was
exhausted or  lost, leaving the  last three surviving  companions Aa2,
Ab, and B.  This  scenario, although speculative, matches the observed
properties of HD~91962.

Most quadruple systems consist of  two close pairs in a 2+2 hierarchy,
while  the  2-tier hierarchies  of  3+1  ``planetary''  type are  less
typical;   they  are   found  in   about  1\%   of   solar-type  stars
\citep{Tok2014}.  The  sample of  4847 solar-type stars  within 67\,pc
contains only  24 multiple  systems with a  3-tier hierarchy,  but for
none of  them except HD~91962 are  all three orbits  known because the
outer orbits have estimated periods of several thousand years.  In the
current
version\footnote{\url{http://www.ctio.noao.edu/\~{}atokovin/stars/index.php}}
of  the  Multiple  Star  Catalog  \citep{MSC}  we  found  four  3-tier
hierarchies  where all three  orbits are  known (Table~\ref{tab:msc}).
In  those  systems,  the  inner  pairs  have  short  orbital  periods,
presumably produced by inward  migration. In the first system, HD~5408
(HR~266,  ADS~784), the  two outer  orbits are  nearly  co-planar with
periods   of   83.1\,yr  and   4.85\,yr   (period   ratio  17.1)   and
eccentricities of 0.24 and 0.22.

The architecture of HD~91962 is therefore rare, but not unique. It may
belong to a class of multiple  systems that evolved in a viscous disk.
The distinguishing features of this class are approximate co-planarity
of the orbits,  moderate period ratio on the order  of 20 (possibly in
resonance),  and small  eccentricities. Other  members of  this class,
quadruple as well as triple, may be found among known multiple systems
and  discovered  in the  future.   Determination  of accurate  orbital
elements will be essential in checking the co-planarity and resonance.

\begin{deluxetable}{ cc lll   }          
\tabletypesize{\scriptsize}     
\tablecaption{Three-tier hierarchies with all known orbits in the MSC
\label{tab:msc}  }
\tablewidth{0pt}                                                                           
\tablehead{  
\colhead{HD} &
\colhead{Sp. type} &
\colhead{Inner} &
\colhead{Middle} &
\colhead{Outer} 
}
\startdata
5408  & B9IVn & SB2 4.24\,d   &  SB1,VB 4.84\,yr & VB 83.1\,yr   \\
9770  & K4V   & Ecl. 0.477\,d & VB 4.56\,yr      & VB 123.5\,yr   \\
12376 & G9V   & SB2 3.08\,d   & VB,SB2 12.9\,yr  & VB 330\,yr     \\
21364 & B9Vn  & SB2 7.15\,d   & SB  145\,d       & VB 212\,yr  
\enddata
\end{deluxetable}

This  is the  case when  a common  visual binary  turns into  a unique
object  worth of further  detailed study.   The rare  quadruple system
HD~91962  gives   interesting  insights  about  its   origin  and,  by
extension,  the  origin of  multiple  stars  in  general.  Further  RV
monitoring will  help to confirm the  sign of the  long-term RV trend,
hence the co-planarity of A,B and Aa,Ab.  Precise speckle measurements
of Aa,Ab with high cadence can be used to infer the orientation of the
inner orbit Aa1,Aa2.  This can  be done even better with long-baseline
interferometers.  Direct  resolution of the  inner pair, for  which we
estimate $\Delta K_{\rm  Aa1,Aa2} \sim 3.8$ mag and  separation on the
order  of 20\,mas,  will be  difficult but  not impossible.   The weak
signatures  of B  and Ab  might be  detectable in  the high-resolution
spectra with a  good SNR.  Such observations can  prove the absence of
additional close  companions in this system and  will provide accurate
measurements of  stellar masses.  Future precise  astrometry with {\it
  Gaia} will add new constraints.

\acknowledgements
The data used in this work were obtained at the Southern Astrophysical
Research  (SOAR)   telescope,  which  is   a  joint  project   of  the
Minist\'{e}rio  da  Ci\^{e}ncia,  Tecnologia,  e  Inova\c{c}\~{a}o  da
Rep\'{u}blica  Federativa  do   Brasil,  the  U.S.   National  Optical
Astronomy  Observatory, the  University  of North  Carolina at  Chapel
Hill, and  Michigan State University. 

This work  used the  SIMBAD service operated  by Centre  des Donn\'ees
Stellaires  (Strasbourg, France),  bibliographic  references from  the
Astrophysics Data System maintained  by SAO/NASA, data products of the
Two  Micron All-Sky  Survey (2MASS),  and the  Washington  Double Star
Catalog maintained at USNO.






\facility{Facility: SOAR}


\begin{thebibliography}{99}

\bibitem[Aitken(1904)]{Aitken1904}
Aitken, R. G. 1904, Lick Obs. Bull. 2, 139,

\bibitem[Buchhave et al.(2014)]{Bucchave2014}
Buchhave, L.A., Bizzaro, M, Latham, D. W., et al. 2014,  Nature, 509, 593.

\bibitem[Carpenter et al.(2009)]{Carpenter2009}
Carpenter, J. M., Bouman, J., Mamajek, E. E. et al.  2009, ApJS, 181, 197

\bibitem[Casagrande et al.(2011)]{Casagrande2011}
Casagrande, L., Schoenrich, R., Asplund, M. et al. 2011 A\&A, 530, 138


\bibitem[Cutri et al.(2003)]{2MASS}
Cutri, R. M., Skrutskie, M. F., van Dyk, S. et al. 2003 The IRSA
2MASS All-Sky Point Source Catalog. NASA/IPAC Infrared
Science Archive.


\bibitem[Dotter et al.(2008)]{Dotter2008}
Dotter, A., Chaboyer, B., Jevremovi\'c, D. et al. 2008, \apjs, 178, 89

\bibitem[Duch\^ene \& Kraus(2013)]{DK13}
Duch\^ene, G. \& Kraus, A. 2013, ARAA, 51 



\bibitem[Fabricius \& Makarov(2000)]{FM2000}
 Fabricius, C. \& Makarov, V. V. 2000, A\&A, 356, 141


\bibitem[Horch et al.(2001)]{Horch2001}
Horch, E. P., Ninkov, Z., \& Franz, O. G., 2001, AJ, 121, 1583


\bibitem[Latham(1992)]{Latham1992}
Latham, D. W. 1992, in ASP Conf. Ser. 32, Complementary Approaches to
Binary and Multiple Star Research, ed. H. McAlister \& W. Hartkopf
(IAU Colloq. 135) (San Francisco: ASP), 110

\bibitem[Latham(1985)]{Latham1985}
Latham, D. W. 1985, in IAU Colloq. 88, Stellar Radial Velocities, ed.
A. G. D. Philip \& D.W. Latham (Schenectady: L. Davis), 21

\bibitem[Metchev \& Hillenbrand(2009)]{MH09}
 Metchev S. A., Hillenbrand L. A. 2009, ApJS, 181, 62

\bibitem[Popovic(1978)]{Pop1978}
 Popovic, G. M. 1978,   Bull. Obs. Astron. Belgrade  No. 129, 9

\bibitem[Schroeder et al.(2009)]{Schroeder}
Schroeder, C., Reiners, A., \& Schmitt, J. H. M. M. 2009, A\&A, 493, 1099 

\bibitem[Szentgyorgyi \& Fur\'esz(2007)]{TRES}  
Szentgyorgyi, A.  H., \&  Fur\'esz, G.  2007,  in The  3rd Mexico-Korea
Conference  on Astrophysics: Telescopes  of the  Future and  San Pedro
M\'artir,   ed.  S.    Kurtz,  RMxAC, 28, 129

\bibitem[Tokovinin(1997)]{MSC}
Tokovinin, A. 1997, A\&AS, 124, 75 


\bibitem[Tokovinin et al.(2010)]{TMH10}
 Tokovinin, A., Mason, B., \& Hartkopf, W. 2010,  AJ, 139, 743

\bibitem[Tokovinin et al.(2014)]{TMH14}
 Tokovinin, A., Mason, B., \& Hartkopf, W. 2014,  AJ, 147, 123


\bibitem[Tokovinin(2014)]{Tok2014}
Tokovinin, A. 2014, AJ, 147, 87

\bibitem[van Leeuwen(2007)] {HIP2}
van Leeuwen, F. 2007, A\&A, 474, 653


\bibitem[White et al.(2007)]{White07}
White, R. J., Gabor, J. M., \& Hillenbrand, L. A. 2007, AJ, 133, 252

\bibitem[Zucker \& Mazeh(1994)]{TODCOR}	
Zucker, S. \& Mazeh, T. 1994, ApJ, 420, 806








\end{thebibliography}
\end{document}